# Baryons in the relativistic jets of the stellar-mass black hole candidate 4U1630-47

Authors: María Díaz Trigo[1], James C.A. Miller-Jones[2], Simone Migliari[3], Jess W. Broderick[4], Tasso Tzioumis[5]

Affiliations:

1. European Southern Observatory, Karl-Schwarzschild-Strasse 2, 85748, Garching bei München, Germany

2. International Centre for Radio Astronomy Research, Curtin University, GPO Box U1987, Perth, WA 6845, Australia

3. Departament d'Astronomia i Meteorologia, Universitat de Barcelona, Martí I Franquès 1, 08028 Barcelona, Spain

4. School of Physics & Astronomy, University of Southampton, SO17 1BJ Southampton, UK

5. Australia Telescope National Facility, CSIRO, PO Box 76, Epping, NSW 1710, Australia

**Accreting black holes are known to power relativistic jets, both in stellar-mass binary systems and at the centres of galaxies. The power carried away by the jets, and hence the feedback they provide to their surroundings, depends strongly on their composition. Jets containing a baryonic component should carry significantly more energy than electron-positron jets. While energetic considerations[1,2] and circular polarisation measurements[3] have provided conflicting circumstantial**

**evidence for the presence or absence of baryons, the only system in which baryons have been unequivocally detected in the jets is the X-ray binary SS 433[4,5]. Here we report the detection of Doppler-shifted X-ray emission lines from a more typical black hole candidate X-ray binary, 4U1630-47, coincident with the reappearance of radio emission from the jets of the source. We argue that these lines arise in a jet with velocity 0.66c, thereby establishing the presence of baryons in the jet. Such baryonic jets are more likely to be powered by the accretion disc[6] rather than the spin of the black hole[7], and if the baryons can be accelerated to relativistic speeds, should be strong sources of gamma rays and neutrino emission.**

As part of a campaign to study the connection between relativistic jets and disc winds in X-ray binaries, we used XMM-Newton and the Australia Telescope Compact Array (ATCA) to make two quasi-simultaneous observations of the stellar-mass black hole candidate X-ray binary 4U1630-47 during its 2012 outburst (see Extended Data Table 1). The first observation, on 2012 September 11-12, showed an X-ray spectrum fully consistent with emission from a standard accretion disc. The quasi-simultaneous radio observation did not detect the source, to 3$\sigma$ upper limits of 68 $\mu$Jy/beam at both 5.5 and 9.0 GHz. In the second observation, on September 28, an additional component was required to model the X-ray spectrum; either a non-thermal power-law, thermal bremsstrahlung or a Comptonisation component (see Extended Data Table 2). Furthermore, three narrow X-ray emission lines were significantly detected at energies of 4.04, 7.28 and 8.14 keV (see Fig. 1).

The quasi-simultaneous ATCA observations detected a radio source at levels of 110±17 and 66±28 μJy beam$^{-1}$ at 5.5 and 9.0 GHz, respectively.

The strongest emission line, at 7.28 keV, is narrow, with a width of 0.17±0.05 keV. There are no known lines with this rest energy and the most plausible explanation is as blueshifted emission from highly-ionised Fe. However, the narrow line width, strong blueshift and the lack of an extended red wing argue against a disc reflection line similar to those previously observed in both X-ray binaries and Active Galactic Nuclei[8]. While narrow emission lines are frequently observed from the corona in very high inclination X-ray binaries (the so-called accretion disc corona sources), they are not significantly blueshifted (see Supplementary Information). Given the reactivation of the jets implied by the onset of radio emission, we are left with the intriguing possibility that the 7.28 keV emission line arises from a relativistic jet moving towards the observer.

Taken in isolation, this would constrain the jet velocity to be >0.3$c$ and the inclination angle to be <73° to the line of sight, assuming Fe XXVI Kα emission (>0.4$c$ and <66°, respectively, for Fe XXV Kα emission). However, if the 4.04 keV line is associated with the corresponding redshifted Fe XXVI Kα emission from the receding jet, we can uniquely constrain the jet velocity to be 0.66$c$, and the inclination angle of the jet axis to the line of sight to be 65° (0.63$c$ and 63°, respectively, for Fe XXV Kα emission). Given that the disc normal is constrained to be in the range 60-75° to the line of sight from the observed X-ray dips and the absence of eclipses[9], and assuming that the outer and inner discs have the same inclination (i.e. there is no warp), this suggests that the jets are aligned with the disc normal, and hence perpendicular to the disc plane. Further, by assuming identical blueshifts for the 7.28 and 8.14 keV lines, we find a unique identification of such lines as Fe XXVI and Ni XXVII and can self-consistently fit the observed emission lines and the hard continuum component as an emission spectrum from hot, diffuse, Doppler-shifted gas at a temperature of $21 \pm 4$ keV, which contributes 19% of the total 2-10 keV unabsorbed flux (see Fig. 2).

Fitting a standard multicolour disc and power law with a photon index of $\Gamma=2$ to model the continuum flux in the second observation gave a bolometric flux in excess of $5.2 \times 10^{-8}$ erg cm$^{-2}$ s$^{-1}$. However, the exact value depends significantly on the relatively poorly constrained photon index. For a photon index of $\Gamma=2$ the power law component comprised 50% of the total bolometric luminosity, while for $\Gamma=2.5$ the power law contribution was as high as 90%. In either case, this implies that the source was in the "anomalous" accretion state[10], characterized by a luminosity in excess of $2.5 \times 10^{38}$ erg s$^{-1}$ with a dominant contribution to the spectrum from a steep power law component. The accretion flow is believed to consist of a standard disc, but with a hot corona responsible for Compton upscattering a significant fraction of the disc photons. The only previous detection of jets from 4U1630-47 also occurred during an "anomalous" state, with the detection of highly polarized, optically thin radio emission[11], although no high-resolution X-ray spectra from this outburst are available to search for Doppler-shifted line emission. Contrasting the accretion flow geometry with that of the standard high/soft state in which jets are quenched by factors of at least several hundred[12] then suggests that it is the presence of the corona, rather than the absence of a standard disc, that is responsible for the launching of the jets.

Narrow, Doppler-shifted X-ray emission lines have previously been reported during an "apparently standard", slim disc state of 4U1630-47[13]. However, the poorer spectral resolution of RXTE compared to XMM-Newton prevented the definitive association of these lines with either the accretion disc or the jet. While we find that interpreting these lines as red- and blueshifted Fe XXVI K$\alpha$ emission from a bipolar jet would give a consistent inclination angle of 58-67$^\circ$ and a slightly lower jet speed of 0.3-0.4$c$, the spectral resolution is insufficient to draw more concrete conclusions. Therefore, until now, the only X-ray binary for which there was unambiguous evidence for baryons in the jets was SS 433, in which Doppler-shifted emission lines are seen in both the optical and X-ray bands[4,5]. However, its persistent, supercritical accretion rate makes it unclear how that system relates to other, more canonical X-ray binaries with significantly more relativistic jets. While baryons could be fed to the jets of SS 433 via its strong stellar wind, this mechanism cannot be invoked in other, lower-mass X-ray binaries.

4U1630-47 is a recurrent transient system. The X-ray spectral and timing features observed during its many well-studied outbursts are typical of other low-mass X-ray binaries, and, together with the absence of Type I X-ray bursts, make it one of the best candidates to contain a black hole[14]. However, the high column density towards the system has precluded spectroscopic classification of the detected infrared counterpart[15], and a dynamical mass estimate is still lacking. The donor star is most likely to be a relatively early-type (late B or F-class) star, similar to 4U1543-47, GRO J1655-40 or SAX J1819.3-2525, with accretion occurring via Roche lobe overflow[15]. One possible explanation for the non-detection of lines from other systems is that relativistic expansion of standard X-ray binary jets would sufficiently Doppler broaden any emission lines to the point that their detection would be rendered extremely difficult, even if one knew the true jet velocity and inclination angle, and hence the expected redshifts[16]. The observed emission lines in 4U1630-47 could then arise from the particular characteristics of the jets during the poorly-studied "anomalous", high-luminosity state in which they were observed.

Most previous attempts to constrain jet composition in both X-ray binaries and AGN have relied on energetics considerations, since baryon-loaded jets can carry significant kinetic power away from the central compact objects without radiating. In some AGN, the detection of circular polarization has been used to determine the low-energy electron population, and hence claim, on energetic grounds, to rule out a significant baryonic component[3]. However, the few reported circular polarization detections in X-ray binaries[17] were unable to place strong constraints on the jet composition. Arguments based on pressure balance and minimum energy calculations in the lobes of radio galaxies have suggested that cold protons may carry the bulk of the kinetic energy[1], but there are caveats to these interpretations and uncertainty in the jet composition remains[18]. In the best-constrained X-ray binary system, calorimetry of the jet-blown bubble around Cygnus X-1 suggested that the jets should carry a significant cold proton component[2], although alternative explanations were possible[19]. The detection of baryons in the jets of 4U 1630-47 has finally confirmed this picture, at least for certain accretion states.

If they can be accelerated to mildly relativistic speeds, the presence of baryons in an X-ray binary jet suggests that gamma-rays could be produced by collisions with high-energy photons or with protons in the stellar wind of the companion star[20,21]. Such hadronic mechanisms are in principle capable of explaining the observed gamma-ray flux from the microquasar Cygnus X-3[22,23], although leptonic models appear equally viable[24]. Even for low-mass X-ray binaries with no strong stellar wind, hadronic models suggest that the presence of relativistic baryons would give rise to gamma-ray emission that could be detected by Fermi-LAT, MAGIC II and the CTA[21]. In that case, the hadronic mechanism should also generate an intense flux of neutrinos[25]. Thus, baryonic jets also have important implications for current and future neutrino telescopes[26], and our results suggest that high-luminosity outbursts could provide the best opportunities for neutrino detection.

Finally, the jet composition should be affected by the physical mechanism responsible for launching the jets. Jets powered by an accretion disc[6] are expected to contain baryons, whereas in the absence of entrainment (as is likely the case in the absence of a strong stellar wind), jets powered by black hole spin[7] are more likely to produce purely leptonic jets with significantly higher Lorentz factors. While there are claims that transient jets from X-ray binaries are powered by black hole spin[27], this work remains controversial[28], so additional evidence detailing the jet composition can provide independent constraints on the jet launching mechanism.

**References:**

1. Dunn, R. J. H., Fabian, A. C. & Taylor G. B. Radio bubbles in clusters of galaxies. *Mon. Not. R. Astron. Soc.* **364**, 1343-1353 (2005).
2. Gallo, E. et al. A dark jet dominates the power output of the stellar black hole Cygnus X-1. *Nature* **436**, 819-821 (2005).
3. Wardle, J. F. C., Homan, D. C., Ojha, R. & Roberts, D. H. Electron-positron jets associated with the quasar 3C279. *Nature* **395**, 457-461 (1998).
4. Margon, B., Grandi, S. A., Stone, R. P. S. & Ford H. C. Enormous periodic Doppler shifts in SS 433. *Astrophys. J.* **233**, L63-L68 (1979).
5. Kotani, A. et al. Discovery of the double Doppler-shifted emission-line systems in the X-ray spectrum of SS 433. *Pub. Astron. Soc. Japan.* **46**, L147-L150 (1994).
6. Blandford, R. D. & Payne, D. G. Hydromagnetic flows from accretion discs and the production of radio jets. *Mon. Not. R. Astron. Soc.* **199**, 883-903 (1982).
7. Blandford, R. D. & Znajek, R. L. Electromagnetic extraction of energy from Kerr black holes. *Mon. Not. R. Astron. Soc.* **179**, 433-456 (1979).
8. Miller, J. M. Relativistic X-Ray Lines from the Inner Accretion Disks Around Black Holes. *Ann. Rev. Astron. Astophys.* **45**, 441-479 (2007).

9. Kuulkers, E. et al. Absorption Dips in the Light Curves of GRO J1655-40 and 4U 1630-47 during Outburst. *Astrophys. J.* **494**, 753-758 (1998).

10. Abe, Y., Fukazawa, Y., Kubota, A., Kasama, D. & Makishima, K. Three Spectral States of the Disk X-Ray Emission of the Black-Hole Candidate 4U 1630- 47. *Pub. Astron. Soc. Japan* **57**, 629-641 (2005).

11. Hjellming, R. M. et al. Radio and X-Ray Observations of the 1998 Outburst of the Recurrent X-Ray Transient 4U 1630-47. *Astrophys. J.* **514**, 383-387 (1999).

12. Russell, D. M. et al. Testing the Jet Quenching Paradigm with an Ultradeep Observation of a Steadily Soft State Black Hole. *Astrophys. J.* **739**, L19 (2011).

13. Cui, W., Chen, W., Zhang, S. N. Evidence for Doppler-shifted Iron Emission Lines in Black Hole Candidate 4U 1630-47. *Astrophys. J.* **529**, 952-960 (2000).

14. McClintock, J. E., Remillard, R. A. in *Compact Stellar X-ray Sources* (eds Lewin, W. H. G. & van der Klis, M.) 157-213 (Cambridge Univ. Press, Cambridge, UK, 2006).

15. Augusteijn, T., Kuulkers, E., van Kerkwijk, M. H. The IR counterpart of the black-hole candidate 4U1630-47. *Astron. Astrophys.* **375**, 447-454 (2001).

16. Mirabel, I. F., Bandyopadhyay, R., Charles, P. A., Shahbaz, T., Rodriguez, L. F. The Superluminal Source GRS 1915+105: A High Mass X-Ray Binary? *Astrophys. J.* **477**, L45-L48 (1997)

17. Fender, R. P. et al. Variable circular polarization associated with relativistic ejections from GRS 1915 + 105. *Mon. Not. R. Astron. Soc.* **336**, 39-46 (2002).

18. Worrall, D. The X-ray jets of active galaxies. *Astron. Astrophys. Review*, **17**, 1-46 (2009).

19. Heinz, S. Composition, collimation, contamination: the jet of Cygnus X-1. *Astrophys. J.* **636**, 316-322 (2006).

20. Romero, G. E., Torres, D. F., Kaufman Bernadó, M. M. & Mirabel, I. F. Hadronic gamma-ray emission from windy microquasars. *Astron. Astrophys.* **410**, L1-L4 (2003).

21. Vila, G. S., Romero, G. E. & Casco, N. A. An inhomogeneous lepto-hadronic model for the radiation of relativistic jets. Application to XTE J1118+480. *Astron. Astrophys.* **538**, A97, 1-12 (2012).

22. Tavani, M. et al. Extreme particle acceleration in the microquasar Cygnus X-3. *Nature* **462**, 620-623 (2009).

23. The Fermi LAT Collaboration, Modulated High-Energy Gamma-Ray Emission from the Microquasar Cygnus X-3. *Science* **326**, 1512-1516 (2009).

24. Piano, G. et al. The AGILE monitoring of Cygnus X-3: transient gamma-ray emission and spectral constraints. *Astron. Astrophys.* **545**, A110, 1-12 (2012).

25. Levinson, A. & Waxman, E. Probing microquasars with TeV neutrinos. Physical Review Letters, **87**, 171101 (2001).

26. Aiello, S. et al. Sensitivity of an underwater Čerenkov km$^3$ telescope to TeV neutrinos from Galactic microquasars. *Astroparticle Phys.* **28**, 1-9 (2007).

27. Narayan, R. & McClintock, J. E. Observational evidence for a correlation between jet power and black hole spin. *Mon. Not. R. Astron. Soc.* **419**, L69-L73 (2012).

28. Russell, D. M., Gallo, E. & Fender, R. P. Observational constraints on the powering mechanism of transient relativistic jets. *Mon. Not. R. Astron. Soc.* **431**, 405-414 (2013).

29. Begelman, M. C., Hatchett, S. P., McKee, C. F., Sarazin, C. L., & Arons, J. Beam models for SS 433. *Astrophys. J.* **238**, 722-730 (1980).

30. Marshall, H. L., Canizares, C. R., & Schulz, N. S. The high resolution X-ray spectrum of SS 433 using the Chandra HETGS. *Astrophys. J.* **564**, 941-952 (2002).

**Supplementary Information** is linked to the online version of the paper at www.nature.com/nature.

**Acknowledgements:**

Based on observations obtained with XMM-Newton, an ESA science mission with instruments and contributions directly funded by ESA member states and the USA (NASA). We thank the XMM-Newton team for the fast scheduling of these observations and the EPIC calibration team for advice. The Australia Telescope Compact Array is part of the Australia Telescope National Facility which is funded by the Commonwealth of Australia for operation as a National Facility managed by CSIRO. This work was supported by the Australian Research Council's *Discovery Projects* funding scheme (J.C.A.M.-J.; project number DP120102393), the Spanish Ministerio de Economía y Competitividad and European Social Funds through a Ramón y Cajal Fellowship (S.M.), and the Spanish Ministerio de Ciencia e Innovación (S.M.; grant AYA2010-21782-C03-01). M.D.T. thanks Anaelle Maury, Andrzej Zdziarski and Chris Done, J.C.A.M.-J. thanks Matt Middleton and S.M. thanks Gustavo Romero, Valenti Bosch-Ramon, Giovanni Miniutti and Sara Motta for discussions.

**Author Contributions:**

M.D.T., S.M. and J.C.A.M.-J. conceived and designed the observing program. M.D.T. and S.M. analysed the XMM observations. J.W.B. and T.T. took the radio observations, which were reduced by J.C.A.M.-J.. J.C.A.M.-J. and M.D.T. wrote the manuscript, with help from S.M.

**Author Information:**

Reprints and permissions information is available at www.nature.com/reprints. The authors declare no competing financial interests. Correspondence should be addressed to M.D.T. (mdiaztri@eso.org).

**Figures:**

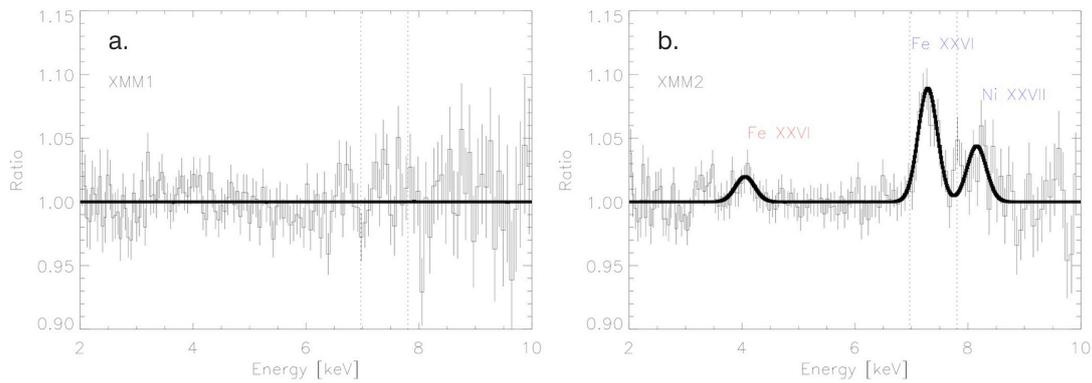

Figure 1: **Residuals from the continuum modelling of the X-ray spectra.** Data (90% error bars) to continuum model ratio. The dotted vertical lines indicate the rest energy of the transitions of Fe XXVI (6.97 keV) and Ni XXVII (7.74 keV). The flux ratio between the blue- and redshifted components of Fe XXVI is from 1.9 ± 1.1 to 2.1 ± 1.3 (see Extended Data Table 3), consistent with 3.2, the ratio predicted for Doppler boosting in a continuous jet. Assuming that the lines are Doppler broadened by divergence in a conical outflow[29,30] we use their widths to determine an upper limit to the opening angle of the jet of 3.7 - 4.5°.

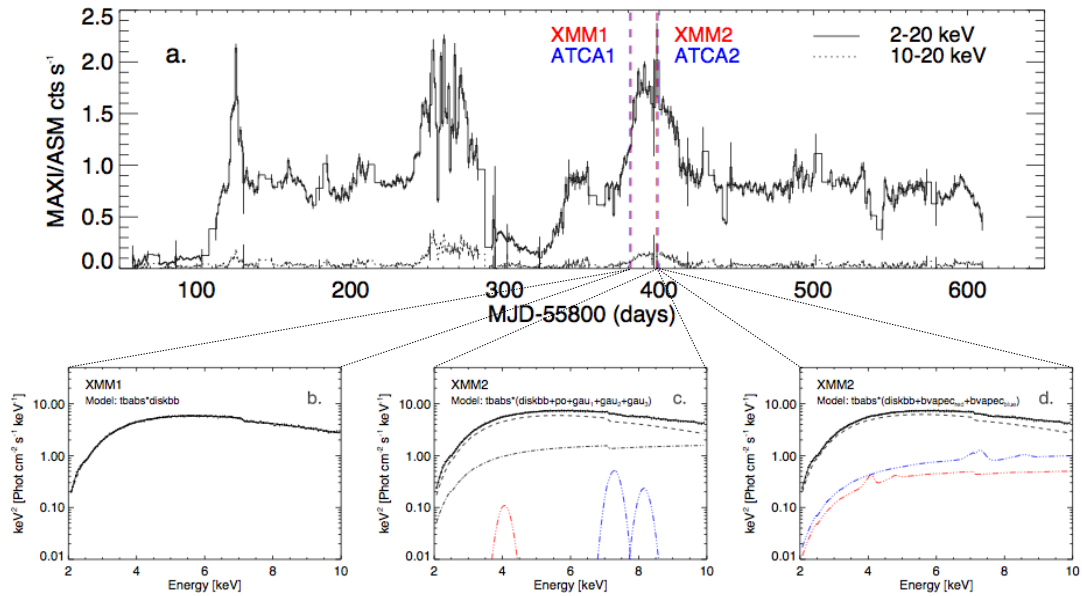

Figure 2: **X-ray observations of 4U1630-47. a.** MAXI/All-Sky-Monitor light curve of 4U 1630-47 (time in days refers to Modified Julian Date, MJD). **b-d** 2-10 keV unfolded X-ray spectra (90% error bars). The first observation can be fully described by a standard disc (**b**). As the source brightens above 2.5 × $10^{38}$ erg s$^{-1}$ in the 2-10 keV band, the spectrum requires an additional hard, power-law, component (black dot-dashed line) and three narrow emission lines (red and blue dot-dashed lines) (**c**). Alternatively, two thermal plasma ("bvapec") components with a temperature of 21 keV (red and blue dot-dashed lines) can account self-consistently for both the hard component and the narrow emission lines (**d**).

## Methods summary

The X-ray observations were made with XMM-Newton, using the EPIC pn CCD camera in burst mode. Data were reduced using the XMM-Newton Science Analysis Software. We corrected for a rate dependent charge transfer inefficiency and rebinned the data, but did not perform background subtraction, owing to both the source brightness and the lack of source-free background regions in burst mode. The resulting X-ray spectra were fit using the spectral analysis package Xspec. While the spectrum from 2012 September 11th could be well fit with a standard disc model, that from September 28th required an additional hard component (Extended Data Table 2). Having modelled the spectral continuum, a strong, narrow emission feature remained in the residuals of the latter observation, at an energy of ~7.3 keV. Additional, weaker features were detected at ~4.1 and ~8.2 keV, whose high statistical significance was confirmed both by an F-test and by Monte Carlo simulations. We also verified the lines to be robust against systematic shifts of up to 2% in the energy scale (although the model became degenerate if the line energy was shifted to coincide with the neutral Fe edge). The best fitting model parameters are given in Extended Data Table 3.

Quasi-simultaneous radio observations were made with the Australia Telescope Compact Array in its compact H214 configuration, using the Compact Array Broadband Backend to observe simultaneously at 5.5 and 9.0 GHz. Data were initially imported into Miriad, and further processed within the Common Astronomy Software Application (CASA), using standard procedures. PKS B1934-638 was used as the amplitude and bandpass calibrator, and

PMN J1603-4904 as the secondary calibrator. No radio emission was detected on 2012 September 10th, with an upper limit of three times the rms noise level. The steep-spectrum emission seen on September 29th was fit with an elliptical Gaussian in the image plane.

## Methods

**Observations and data analysis:**

*XMM-Newton:*

In this section we provide additional information about the analysis of the XMM-Newton observations of 4U 1630-47. A summary of the observations is shown in Extended Data Table 1. Due to the high count rate from the source, $\geq 1000$ cts s$^{-1}$, we used the EPIC pn CCD camera in burst mode, in which only one CCD chip is operated and the data are collapsed into a one-dimensional row (4'.4) and read out at high speed, the second dimension being replaced by timing information, and applied the Science Analysis Software (SAS) task epfast on the event files to correct for a Charge Transfer Inefficiency (CTI) effect which has been observed in this mode when high count rates are present[31]. We did not use the EPIC MOS CCD cameras during the observations to avoid telemetry overflows. Data products were reduced using the XMM-Newton SAS version 12.0.1. We rebinned the EPIC pn spectra to over-sample the full width at half maximum (FWHM) of the energy resolution by a factor of 3 and to have a minimum of 25 counts per bin, to allow the use of the $\chi^2$ statistic. To account for systematic effects, we added a 0.8% uncertainty to each spectral bin after rebinning. We used the pn spectra between 2 and 10 keV.

In the EPIC pn burst mode there are no source-free background regions, since the point spread function (PSF) of the telescope extends further than the central CCD boundaries[32,33]. Since 4U1630−47 is very bright, its spectrum will not be significantly modified by the "real" background, which contributes less than 1% to the total count rate across most of the energy band. Conversely, subtracting the background extracted from the outer columns of the central CCD will modify the source spectrum, since the PSF is energy-dependent and the source photons scattered to the outer columns do not show the same energy dependence as the photons focused on the inner columns. Therefore, we checked the effect of subtracting the "background" extracted from the outer regions of the central CCD on the parameters of the lines. Having established that this resulted in no significant changes to the lines, we chose not to subtract such a "background" when making the final fits, to provide the best possible measurement of the true source spectrum.

We fitted the XMM-Newton data using the spectral analysis package, Xspec[34], testing standard X-ray binary models for single disc emission (diskbb in Xspec), power-law emission (po in Xspec), bremsstrahlung (bremss in Xspec), thermal Comptonisation (comptt in Xspec), disc emission plus power-law (diskbb+po), disc emission plus bremsstrahlung (diskbb+bremss) and disc emission plus thermal Comptonisation (diskbb+comptt). In each case we included a neutral absorber (tbabs in Xspec) to account for interstellar absorption along the line of sight. We also included a narrow emission feature at 2.28 keV to account for residual calibration uncertainties at the gold edge of the pn camera[31].

We found that the first observation could be well fitted by a standard disc (reduced $\chi_\nu^2$ = 0.97 for 128 degrees of freedom). In contrast, for the second observation, either a one-component thermal Comptonisation model or a two-component model was required to obtain an acceptable fit (reduced $\chi_\nu^2 \leq 2$). A summary of the models tested and the corresponding quality of the fits is shown in Extended Data Table 2. The main difference between the first and the second observation is the appearance of a hard component in the latter observation in addition to the thermal disc component. Further support for the existence of this component is given by the increase of the hard X-ray flux ($\geq$10 keV) between the first and the second observation, as detected by MAXI/All-Sky-Monitor and Swift/BAT. In particular, the SWIFT/BAT 15-50 keV count rate increased from 0.019 ± 0.001 cts cm$^{-2}$ s$^{-1}$ to 0.036 ± 0.002 cts cm$^{-2}$ s$^{-1}$. However, we cannot discriminate between the three components (power-law, bremsstrahlung or thermal Comptonisation) used to model the additional emission in the second observation due to the low effective area of XMM-Newton at energies > 10 keV, at which such a component should become dominant. We found that a one-component, thermal Comptonisation model with low seed photon and plasma temperature (~0.7 keV and 1.9 keV, respectively) and a high optical depth ($\tau \sim 8$) could fit the second observation equally well as the two-component models. However, the 15-50 keV flux predicted by the one-component model, ~3 x 10$^{-10}$ erg cm$^{-2}$ s$^{-1}$, is lower than the flux detected by Swift/BAT, ~2 x 10$^{-9}$ erg cm$^{-2}$ s$^{-1}$, by almost one order of magnitude. In contrast, the two-component models predict a 15-50 keV flux of ~2-3 x $10^{-9}$ erg cm$^{-2}$ s$^{-1}$, in excellent agreement with Swift/BAT. Therefore we discard the one-component thermal Comptonisation model for the rest of this work, but include it for completeness in Extended Data Tables 2 and 3, together with all the other tested models.

Having modelled the continuum emission, a strong narrow emission feature remained in the residuals of the second observation at ~7.3 keV, with additional weaker features at energies of ~ 3.5, 4.1, 7.8 and 8.2 keV. Inclusion of a narrow Gaussian at ~ 7.28 keV to account for the most significant feature improved the fit from $\chi_v^2$ = 1.70 (127 d.o.f.) to 1.07 (124 d.o.f.). The probability for such an improvement occurring by chance, as indicated by an F-test, is 5 x $10^{-13}$. We also tested the significance of the weaker features including them one by one in addition to the feature at 7.28 keV. Since the weak features are narrow, we coupled their widths to that of the 7.28 keV feature to prevent an arbitrary increase of the line widths from absorbing deficiencies in the continuum model. We found that independently including the ~4.1 keV and ~8.2 keV features improved the quality of the fit to $\chi^2$ = 0.97 (122 d.o.f.) and 0.95 (122 d.o.f.), respectively (an F-test suggesting that the probability for such an improvement occurring by chance was 0.0007 and 0.0002, respectively). For the feature at 3.5 keV, the chance probability of the fit improvement as given by the F-test was 0.023. The feature at 7.8 keV is weaker than that at 8.2 keV, and unless the energy was fixed, the fit shifted the energy to 8.2 keV. If the feature at 7.8 keV was included in addition to the

8.2 keV feature, the chance probability of the fit improvement as given by the F-test was 0.041.

Since there are caveats to the use of the F-test to study the presence of emission lines over a continuum[35], we also performed Monte Carlo simulations in order to confirm the significance of the 4.1 keV and the 8.2 keV features[36]. As an example, we took the best-fitting parameters of the model consisting of absorbed disc emission, thermal Comptonisation (diskbb+comptt) and the 7.28 keV Gaussian emission line as our null hypothesis. We then tested the chance probability of any extra emission line against the null hypothesis.

For this, we fitted the data with the null hypothesis model and simulated a spectrum with the same exposure time as the data. We fitted the simulated spectrum with the model used to construct it, providing a refined null hypothesis model that differed from the previous one only by photon statistics. We then added to the model a Gaussian emission line, varying its energy from 3.0 keV to 9.0 keV in steps of 0.2 keV. We chose the line width to be equal to the one that we found in our best-fit model of the data with the three Gaussian emission lines, $\sigma = 0.17 \pm 0.05$ keV, and we allowed the normalisation to vary. For each spectrum, we recorded the best-fitting parameters and the maximum $\Delta\chi^2$ among the lines. We then repeated the above steps 1000 times and obtained the distribution of maximum $\Delta\chi^2$ to be compared with the result obtained from the data. The addition of an extra

Gaussian to the data at 4.1 keV or at 8.2 keV improved the fit by $\Delta\chi^2 = 12.73$ and $\Delta\chi^2 = 13.77$, respectively, compared to the null hypothesis.

Only six and one of the 1000 simulated spectra showed a maximum $\Delta\chi^2$ equal to or higher than those found fitting the data with the additional 4.1 and 8.2 keV lines, respectively. Therefore, these additional lines are significant at levels of 99.4% and 99.9%, respectively.

In summary, given the higher significance of the ~4.1 keV and ~8.2 keV features, we consider them as likely to be real, and discard the other weak features, which have significantly higher chance probabilities. In Extended Data Table 3 we show the parameters of the best fitting models. While the lines are robust against significance tests, we also tested whether they are robust against possible systematic errors in the energy scale. For this, we shifted the energy scale by up to ± 2% in the events files of both observations and re-extracted spectra and response matrices. Coincidentally, the maximum possible systematic error of -2% in the energy scale would shift the photons of the line at 7.28 keV to the energy of the neutral Fe edge causing model degeneracy. Therefore we encourage future exposures with high resolution gratings to mitigate against possible systematics and definitively confirm the exact energies of the lines.

Finally, we searched for line variability within the second observation by dividing the events file into two, three and four different parts and extracting and fitting the corresponding spectra. We did not detect any significant variation of the lines among the different intervals. However, due to the poor statistics we obtain large errors and cannot exclude variability with high significance. Longer exposures with high resolution gratings would be required to definitively address the existence of line variability, as might be expected in the case of jet precession.

*Australia Telescope Compact Array:*

The Australia Telescope Compact Array (ATCA) was used to observe 4U1630-47 at two epochs, as detailed in Extended Data Table 1. We made simultaneous observations at 5.5 and 9.0 GHz using the Compact Array Broadband Backend (CABB) system[37]. We observed with 2048 MHz of contiguous bandwidth in each of the two frequency bands, split into 1-MHz channels. The array was in the hybrid H214 configuration, with five of the six antennas within 250 m, and the sixth antenna providing longer (4.5 km) baselines.

Data were imported into Miriad[38], and immediately written out to FITS format for further processing within the Common Astronomy Software Application (CASA[39]). We edited out narrow-band radio frequency interference, before deriving the external gain calibration, using PKS B1934-638 as the bandpass calibrator and to set the amplitude scale. We used PMN J1603-4904 to derive the amplitude and phase gains that were subsequently applied to the target field. To filter out the diffuse emission in the field and enhance the sensitivity to point sources, we only used the long baselines to antenna 6 in making the radio images. We used natural weighting to maximize the image sensitivity. The source was too faint for self-calibration. On 2012 September 10, the source was not detected at either frequency, and we estimated the 3σ upper limit on its flux density as three times the r.m.s. noise in the image. On 2012 September 29, when the source was significantly detected, we measured its flux density by fitting an elliptical Gaussian to the source in the image plane.

**Supplementary references**

31. Guainazzi, M. et al. Evaluation of the spectral calibration accuracy in EPIC-pn fast modes (2012). Available at http://xmm.esac.esa.int/

32. Done, C. & Díaz Trigo, M. A re-analysis of the iron line in the XMM-Newton data from the low/hard state in GX 339-4, MNRAS **407**, 2287 (2010)

33. Ng, C., Díaz Trigo, M., Cadolle Bel, M., & Migliari, S. A systematic analysis of the broad iron Kα line in neutron-star LMXBs with XMM-Newton, A & A **522**, id.A96 (2010)

34. Arnaud, K.A. Astronomical Data Analysis Software and Systems V, eds. Jacoby G. and Barnes J., **101**,17(1996)

35. Protassov, R., van Dyk, D. A., Connors, A., Kashyap, V. L., & Siemiginowska, A. Statistics, Handle with Care: Detecting Multiple Model Components with the Likelihood Ratio Test. *Astrophys. J.* **571**, 545-559 (2002)

36. Miniutti, G. & Fabian, A. C. Discovery of a relativistic Fe line in PG 1425+267 with XMM-Newton and study of its short time-scale variability. *Mon. Not. R. Astron. Soc.* **366**, 115-124 (2006)

37. Wilson, W. E. et al. The Australia Telescope Compact Array Broad-band Backend: description and first results. *Mon. Not. R. Astron. Soc.* **416**, 832-856 (2011).

38. Sault, R. J., Teuben, P. J. & Wright, M. C. H. A Retrospective View of Miriad. *In Shaw, R. A., Payne, H. E., & Hayes, J. J. E., Eds., Astronomical Data Analysis Software and Systems IV, ASP Conf. Ser.* **77**, 433-436 (1995).

39. McMullin, J. P., Waters, B., Schiebel, D., Young, W., & Golap, K. CASA Architecture and Applications. *In Shaw, R. A., Hill, F., & Bell, D. J., Eds., Astronomical Data Analysis Software and Systems XVI, ASP Conf. Ser.* **376**, 127-130 (2007).

**Tables:**

Extended Data Table 1: Observation log

| Observation | Observation ID | Observation Times (UTC) (day.month.year hr:min) |
|---|---|---|
| XMM1 | 0670673101 | 11.09.2012 20:14 – 12.09.2012 05:39 |
| ATCA1 | C2514 | 11.09.2012 01:08 – 11.09.2012 09:03 |
| XMM2 | 0670673201 | 28.09.2012 06:33 – 28.09.2012 21:50 |
| ATCA2 | C2514 | 29.09.2012 02:36 – 29.09.2012 09:58 |

Extended Data Table 2: Quality of the spectral fits ($\chi_\nu^2$) to the XMM-Newton observations, using different continuum models.

| Model | $\chi_\nu^2$ (d.o.f.) | |
| --- | --- | --- |
| | XMM1 | XMM2 |
| tbabs*diskbb | 0.97 (128) | 2.12 (128) |
| tbabs*po | 9.44 (128) | 12.9 (128) |
| tbabs*bremss | 2.59 (128) | 3.26 (128) |
| tbabs*comptt | 0.87 (126) | 1.26 (126) |
| tbabs*(diskbb+po) | 0.95 (127) | 1.70 (127) |
| tbabs*(diskbb+bremss) | 0.96 (126) | 1.70 (126) |
| tbabs*(diskbb+comptt) | 0.95 (127) | 1.67 (127) |

Extended Data Table 3: Parameters for each of the best-fitting models to the 2-10 keV EPIC pn spectra from the XMM-Newton observations (uncertainties are quoted at the 90% level). $N_{H\,abs}$ is the column density of the neutral absorber in units of $10^{22}$ cm$^{-2}$. $k_{dbb}$, $k_{po}$, $k_{bremss}$ and $k_{comptt}$ are the normalizations of the disc blackbody, power-law, bremsstrahlung and Comptonisation components in Xspec units. $kT_{dbb}$, $kT_{bremss}$ and $kT_{bvapec}$ are the temperatures of the disc blackbody, bremsstrahlung and bvapec components in units of keV. For the comptt component, we imposed a lower limit to the temperature of the plasma of 50 keV and an upper limit to the opacity of 2, and coupled the temperature of the seed photons to the temperature of the disc blackbody in the two-component model. The photon index of the power-law, $\Gamma$, has been fixed to 2, since it is poorly constrained.

k$_{blue(red)}$ represents the normalization of the bvapec component in Xspec units. z$_{blue(red)}$ are the velocity shifts for the blue and red bvapec components, respectively. "Ni" represents the nickel abundance with respect to solar abundances for the bvapec components and has been linked for both components. E$_{gau}$, σ, EW and Flux$_{gau}$ represent the energy (in units of keV), width (in units of keV), equivalent width (in units of eV) and 2-10 keV unabsorbed flux (in units of $10^{-11}$ erg cm$^{-2}$ s$^{-1}$) of the Gaussian features, respectively. The width, σ, was tied for all the emission lines during the fit. 'p' indicates that a parameter pegged at its lower/upper limit.

| Obs. | | | | Best-fit model | | | | | | $\chi^2_\nu$ (d.o.f.) |
|---|---|---|---|---|---|---|---|---|---|---|
| | | | | tbabs*diskbb | | | | | | |
| | $N_{Habs}$ | $kT_{dbb}$ (keV) | $k_{dbb}$ | | | | | | | |
| XMM1 | 8.34±0.08 | 1.82±0.01 | 107±3 | – | – | – | – | – | – | 0.97 (128) |
| | | | | tbabs*(comptt+gau+gau+gau) | | | | | | |
| | $N_{Habs}$ | $kT_{bb}$ (keV) | $kT_e$ (keV) | τ | $k_{comptt}$ | $E_{gau}$ (keV) | σ (keV) | Flux$_{gau}$ | EW (eV) | |
| XMM1 | 7.5±0.4 | 0.69$^{+0.08}_{-0.06}$ | 1.63±0.07 | 9.1±0.9 | 3.0±0.3 | – | – | – | – | 0.87 (126) |
| XMM2 | 7.36$^{+0.09}_{-0.26}$ | 0.72$^{+0.05}_{-0.02}$ | 1.86$^{+0.06}_{-0.04}$ | 7.7±0.5 | 3.23$^{+0.09}_{-0.22}$ | 4.06±0.07 | (t) | 1.9±1.0 | 5±2 | 0.79 (119) |
| | | | | | | 7.24±0.03 | 0.11±0.03 | 3.8±0.8 | 23±4 | |
| | | | | | | 8.22±0.10 | (t) | 0.9±0.7 | 7±4 | |
| | | | | tbabs*(diskbb+po+gau+gau+gau) | | | | | | |
| | $N_{Habs}$ | $kT_{dbb}$ (keV) | $k_{dbb}$ | Γ | $k_{po}$ | $E_{gau}$ (keV) | σ (keV) | Flux$_{gau}$ | EW (eV) | |
| XMM1 | 8.5±0.2 | 1.78±0.03 | 113±6 | 2 (f) | <0.6 | – | – | – | – | 0.95 (127) |
| XMM2 | 8.6±0.1 | 1.77±0.03 | 123±7 | 2 (f) | 1.7±0.3 | 4.04$^{+0.10}_{-0.12}$ | (t) | 3.0±1.4 | 8±4 | 0.87 (120) |
| | | | | | | 7.28±0.04 | 0.17±0.05 | 6.2±1.0 | 37±9 | |
| | | | | | | 8.14$^{+0.14}_{-0.18}$ | (t) | 2.5±1.0 | 18±10 | |
| | | | | tbabs*(diskbb+bremsstrahlung+gau+gau+gau) | | | | | | |
| | $N_{Habs}$ | $kT_{dbb}$ (keV) | $k_{dbb}$ | $kT_{brems}$ (keV) | $k_{brems}$ | $E_{gau}$ (keV) | σ (keV) | Flux$_{gau}$ | EW (eV) | |
| XMM1 | 8.6±0.4 | 1.79±0.06 | 102$^{+16}_{-24}$ | >1.2 | 0.9$^{+1.8}_{-0.8}$ | – | – | – | – | 0.95 (126) |
| XMM2 | 8.4$^{+0.4}_{-0.2}$ | 1.74±0.05 | 134$^{+15}_{-25}$ | >5.3 | 1.10$^{+1.74}_{-0.05}$ | 4.04$^{+0.10}_{-0.12}$ | (t) | 3.2±1.3 | 8±4 | 0.87 (119) |
| | | | | | | 7.27±0.04 | 0.17±0.05 | 6.1±0.9 | 37±8 | |
| | | | | | | 8.15$^{+0.15}_{-0.18}$ | (t) | 2.3±0.9 | 18$^{+11}_{-9}$ | |
| | | | | tbabs*(diskbb+comptt+gau+gau+gau) | | | | | | |
| | $N_{Habs}$ | $kT_{dbb}$ (keV) | $k_{dbb}$ | $kT_e$ (keV) | τ | $k_{comptt}$ | $E_{gau}$ (keV) | σ (keV) | Flux$_{gau}$ | EW (eV) |
| XMM1 | 8.5±0.1 | 1.7$^{+0.1}_{-0.3}$ | 132$^{+34}_{-26}$ | 68$^{+432p}_{-18p}$ | <0.58 | <0.75 | – | – | – | – | 0.96 (125) |
| XMM2 | 8.3±0.1 | 1.6±0.2 | 212$^{+60}_{-70}$ | 145$^{+355p}_{-95p}$ | <0.50 | 0.005$^{+0.3}_{-0.004}$ | 4.03$^{+0.11}_{-0.13}$ | (t) | 2.9±1.4 | 8±4 | 0.88 (118) |
| | | | | | | | 7.27±0.04 | 0.17±0.05 | 6.0±1.0 | 37±8 | |
| | | | | | | | 8.15$^{+0.14}_{-0.18}$ | (t) | 2.4±1.0 | 19±11 | |
| | | | | tbabs*(diskbb+bvapec$_{red}$+bvapec$_{blue}$) | | | | | | |
| | $N_{Habs}$ | $kT_{dbb}$ (keV) | $k_{dbb}$ | $kT_{bvapec}$ (keV) | $z_{bvapec,red}$ | $k_{red}$ | $z_{bvapec,blue}$ | $k_{blue}$ | Ni | v (km/s) |
| XMM2 | 8.3±0.1 | 1.77±0.04 | 130±10 | 21±4 | 0.706±0.026 | 2.0±0.8 | -0.044$^{+0.003}_{-0.010}$ | 1.3±0.3 | <5.4 | 5260$^{+4850}_{-1730}$ | 1.04 (121) |

## Supplementary information

**Alternative scenarios for the origin of the lines:**

In the main text, we attributed the X-ray lines to Doppler-shifted lines in a relativistically-moving jet. Here we consider alternative scenarios for the origin of the lines, namely a highly ionised wind and reflection from the disc. We demonstrate that these possibilities can both be ruled out.

4U 1630-47 is known to show a highly ionised disc wind[40]. Due to the inclination of the source, 60-75° [9], the photoionised wind is observed via highly ionised absorption lines against the bright central continuum. For sources at a higher inclination of > 80°, the so-called "accretion disc corona" sources, obscuration of the central continuum by the rim of the disc enables the detection of the photoionised atmosphere or wind via narrow emission lines, which are not visible at lower inclinations due to the brighter continuum emission. As in previous work[40], we also detected highly-ionised absorption lines from Fe XXV and Fe XXVI in one XMM-Newton observation on the 9th of September 2012, performed only two days before the first observation reported here. This implies that a significant change in the geometry of the system between the 9th and the 28th of September would have to be invoked in order to attribute the emission lines reported here to a disc wind. However, a change in the inclination of the system (or the disc, e.g. via precession) with respect to the observer is unlikely, since X-ray observations of the source in 2007 span more than 21 days and all show strong absorption features[40]. In addition, the two observations reported here show a bright, unobscured disc, inconsistent with the condition of obscuration of the central continuum in accretion disc corona systems. Therefore, we conclude that the narrow emission lines cannot originate in a disc wind.

A second origin of line emission in X-ray binaries could be reflection from the disc. However, the width and blueshift of the lines reported here argues against this interpretation. Reflected emission of highly ionised material is expected to show a significant broadening of the lines due to Comptonisation effects. Additional broadening could arise due to relativistic effects if the lines are emitted close to the black hole. In contrast, the observed lines are narrow. Only reflection from cold material in the outskirts of the disc could produce narrow disc lines. However, such lines would not have a sufficient blueshift to account for the line at 7.28 keV. Thus, we also discard the possibility that the lines originate via disc reflection.

Finally, we note that no spectral lines are known at the rest energies of 7.28 or ~8.2 keV. While the line at ~4.1 keV could have been identified with emission from Ca XX at its rest energy, a high overabundance of Ca with respect to other elements like Fe would have to be invoked to explain its presence in the spectrum. However, such an overabundance would certainly have been detected in previous spectra showing a disc wind. In contrast, no absorption lines from any Ca ions were detected in such spectra[40]. Therefore, we do not find any interpretation for the lines unless they have significant velocity shifts.

**Supplementary references**

40. Kubota, A. et al. Suzaku Discovery of Iron Absorption Lines in Outburst Spectra of the X-Ray Transient 4U 1630-472. *Publ. Astron. Soc. Japan* **59**, 185-198 (2007).